\documentclass[referee]{aa}     % LaTeX A&A  Version 3
\usepackage{epsf}
\usepackage{amssymb}
\usepackage{amsmath}
\usepackage{amsxtra}
\usepackage{times}
\usepackage{graphicx}

\begin{document}

   \title{Determination of the properties of the central engine in 
	microlensed QSOs}

   \author{Luis J. Goicoechea$^1$, 
	David Alcalde$^2$, 
	Evencio Mediavilla$^2$,
	Jos\'e A. Mu\~noz$^2$}
 
   \institute{$^1$ Departamento de F\'{\i}sica Moderna, 
		Universidad de Cantabria, 
		Avda. de Los Castros s/n, E-39005 Santander, Spain;
		\email{goicol@unican.es} \\
	$^2$ Instituto de Astrof\'{\i}sica de Canarias, 
	C/ V\'{\i}a L\'actea s/n, E-38200 La Laguna, Spain; 
	\email{dalcalde@ll.iac.es, emg@ll.iac.es, jmunoz@ll.iac.es}}
		  
   \offprints{L. J. Goicoechea}
 
\date{Received date; accepted date}
\date{Submitted: July 2002}
 
% Running title:
\titlerunning{The central engine in microlensed QSOs}
\authorrunning{Goicoechea et al.}
 
\abstract{
We study a recently observed gravitational microlensing peak in the $V$-band
light curve of Q2237+0305A using a relatively simple, but highly consistent 
with the data (the best-fit reduced $\chi^{2}$ is very close to 1), physical 
model. The source quasar is assumed to be a Newtonian geometrically-thin and 
optically-thick accretion disk. The disk has an arbitrary orientation, and
both blackbody and greybody emission spectra are considered. When the
electron-photon scattering plays a role, the greybody spectrum will be a
simplified version of the exact one. In our model the microlensing variability 
result from the source crossing a caustic straight line. The main goal is to
estimate the black hole mass and the mass accretion rate in QSO 2237+0305 as
well as to discuss the power and the weakness of the technique, some possible
improvements, and the future prospects from multifrequency monitoring of new
microlensing peaks. We also put into perspective the new 
methodology and the results on the central engine in QSO 2237+0305. From the
fitted microlensing parameters and reasonable dynamical/cosmological 
constraints, it is concluded that QSO 2237+0305 harbours a central massive black 
hole: 10$^7$ $M_{\odot}$ $< M <$ 6 10$^8$ $M_{\odot}$. While the information
about the central dark mass is very interesting, the mass accretion rate is not 
so well constrained. The typical values of the disk luminosity/Eddington 
luminosity ratio are in the $(1-20)\varepsilon$ range, where $\varepsilon \leq$ 
1 is the emissivity relative to a blackbody and the highest $L/L_{Edd}$ ratio 
corresponds to the largest deflector motion. Therefore, in order to verify 
$L/L_{Edd} \leq$ 1, a relatively small projected peculiar motion of the lens 
galaxy and a greybody emission seem to be favored.  
}

\maketitle

\section{Introduction}

The optical continuum from a QSO could be originated from a geometrically-thin
and optically-thick standard accretion disk. This standard scenario is a good 
candidate to explain most of 
the non-variable background component, whereas the fluctuations on 
different time scales may be caused by different mechanisms (accretion disk 
instabilities, supernova explosions in a circumnuclear stellar region, and so 
on). The Newtonian model of a standard accretion disk around a black 
hole was introduced by Shakura \& Sunyaev (1973). In this model, the released 
gravitational energy is emitted as a multitemperature blackbody radiation, 
where $T_s(r)$ is the temperature at radius $r$. Thus, in the absence of 
extinction, the emitted and observed intensities obey Planck laws 
$B_{\nu_s}[T_s(r)]$ and 
$B_{\nu}[T(r)]$, respectively, where the emitted temperature and the observed 
one are related through the cosmological redshift of the source $z_s$: $T(r) = 
T_s(r)/(1 + z_s)$. As the emitted temperature profile depends on the black hole
mass and the mass accretion rate, the actually observed intensity profile 
$I_{\nu}(r) = \epsilon_{\nu} B_{\nu}[T(r)]$ is determined from the two physical
parameters of the source. In a general situation we must take into account an
extinction factor $\epsilon_{\nu}$ due to dust in the host galaxy of the 
quasar and any galaxy in the way to the observer including a possible lens 
galaxy and the Milky Way. A relativistic version of the Shakura \& Sunyaev 
model was presented by Novikov \& Thorne (1973) and Page \& Thorne (1974). The 
relativistic model should be a useful tool to describe the physics of the 
innermost layers of the disk. However, as we are interested in the emissivity 
of a wide optical source from an inner edge of several Schwarzschild radii up 
to an outer edge of 10$^2$--10$^3$ Schwarzschild radii, in a first approach, we
can ignore the relativistic effects and take the Newtonian picture. 

When a lensed quasar crosses a microcaustic, it is seen an important 
fluctuation in the flux of one of its images, i.e., a gravitational 
microlensing high-magnification event (HME) appears. The microlensing light 
curve of the involved image is basically given by convolving the observed 
intensity distribution with the corresponding magnification pattern. Therefore,
the behaviour of the HME depends on the physical properties of the source
quasar, and this fact makes way to a discussion on the structure of the source 
from the analysis of the prominent microlensing event (e.g., Rauch \& Blandford
1991; Jaroszy\'nski, Wambsganss \& Paczy\'nski 1992). In principle, if we 
assume a standard accretion disk as the source of the optical continuum, a 
measurement of the black hole mass and the mass accretion rate may be possible
and, recently, several papers pointed that the study of an individual HME may be
used to infer these parameters (Yonehara et al. 1998; Agol \& Krolik 1999; 
Mineshige \& Yonehara 1999; Yonehara et al. 1999; Shalyapin 2001; Yonehara 
2001).

Very recently the GLITP (Gravitational Lenses International Time Project) 
collaboration has monitored the four images of QSO 2237+0305 in the $V$ and
$R$ bands. In each optical band, the GLITP light curve for the brightest image,
Q2237+0305A, traced the peak of one HME with an unprecedented quality (Alcalde 
et al. 2002). The global flat shape for the light curve of the faintest image, 
Q2237+0305D, suggested that the intrinsic signal is globally stationary, and
this result supported that the global variability in Q2237+0305A is exclusively 
caused by microlensing. The peak of the HME was fitted to the microlensing
curves resulting from face-on circularly-symmetric sources crossing a single
straight fold caustic, and it was obtained that the only source models totally
consistent with the GLITP data are the Newtonian standard accretion disk and
its simplified versions (Shalyapin et al. 2002). The uniform and Gaussian disks 
led to an excessively high value of the reduced chi-square, but the simplified 
variants of the Newtonian standard accretion disk gave excellent results: the 
best fits are in very good agreement with the observations and the upper limits 
on the source size are highly reasonable. To avoid a possible anomalous ratio of
fitted background fluxes $F_{0R}/F_{0V}$, it could be needful to consider 
additional $R$ light, which is emitted from an extended region (e.g., Jaroszy\'nski, 
Wambsganss \& Paczy\'nski 1992). This possible $R$-band extended source is 
however irrelevant in the estimation of the compact source size as well as in
the measurements of the central mass and the accretion rate from the technique
presented in this paper. Unfortunately, from the framework used by 
Shalyapin et al. (2002), we cannot measure the physical parameters of the 
engine. One can only determine a relationship between the black hole mass 
($M$), the mass accretion rate ($dM/dt$), and the quasar velocity perpendicular
to the caustic line ($V_{\perp}$). Even from some reasonable velocity range, 
there is no way to separately infer $M$ and $dM/dt$. 

In this paper (Section 2), we introduce a novel expression of the microlensing 
light curve when a Newtonian standard source crosses a single straight fold 
caustic. This novel approach permits to break the degeneracy in the estimation 
of $M$ and $dM/dt$, and so, to measure these parameters for a source with an 
arbitary orientation. In Section 3, from the $V$-band microlensing peak found 
by the GLITP collaboration, we obtain estimates of the black hole mass and the 
mass accretion rate in QSO 2237+0305. In Section 4 we present the main conclusions
and put into perspective the technique and the new results on the central engine 
in QSO 2237+0305.                           

\section{Microlensing light curve for a Newtonian standard accretion disk near
to a caustic straight line}

Yonehara et al. (1998) and Shalyapin et al. (2002) have previously studied the
microlensing light curve associated with a Newtonian standard source that
crosses a single straight fold caustic. Yonehara et al. (1998) presented the
expected light curve when a face-on source ($M = 10^8 M_{\odot}$, $dM/dt
\approx 0.13 \hspace{0.07in} M_{\odot}$ yr$^{-1}$) is strongly magnified by a 
given caustic line, while Shalyapin et al. (2002) discussed the time evolution 
of the monochromatic flux for a face-on source (the inner and outer edges were 
assumed to be $r_{in} = 0$ and $r_{out} = \infty$, respectively) near to a 
generic caustic line. Using the theoretical light curve reported by Shalyapin
et al. and an observed HME, one is able to obtain information on the size of 
the involved source quasar, provided that some interval of the quasar velocity 
$V_{\perp}$ might be inferred from observational data. However, their approach 
does not permit a measurement of the two main parameters of the source: the 
central mass and the mass accretion rate.

Here we like to show the behaviour of a HME caused by a generic Newtonian
standard accretion disk (with finite inner and outer edges, and an arbitrary
orientation) crossing a generic caustic straight line, and more importantly, a 
framework in which is possible to measure the physical properties of the black
hole--accretion disk complex. We begin with a basic ingredient of the
monochromatic radiation flux: the observed intensity profile. The observed
intensity from a part of the disk at ($r$,$\psi$) is given by
\begin{equation} 
I_{\nu}(r) = \epsilon_{\nu} B_{\nu}[T(r)] = \frac{2h\nu^3}{c^2}  
\frac{\epsilon_{\nu}}{\exp [h\nu/kT(r)] - 1} ,
\end{equation}
where (Shakura \& Sunyaev 1973)
\begin{equation} 
T(r) = \left[\frac{3}{8\pi} \frac{GM}{\sigma r^3} \left(\frac{dM}{dt}\right) 
\left(1 - \sqrt{\frac{r_{in}}{r}}\right)\right]^{1/4} (1 + z_s)^{-1} .
\end{equation}
The temperature profile depends on the black hole mass ($M$) and the mass
accretion rate ($dM/dt$), as well as the redshift of the source ($z_s$) and the
inner edge of the disk ($r_{in}$). Two physical constants are also involved:
the Stefan constant ($\sigma$) and the gravitation constant ($G$). As usual,
$r_{in}$ is assumed to be thrice the Schwarzschild radius of the black hole,
i.e., $r_{in} = 6 GM/c^2$. Both the temperature and the intensity are
circularly symmetric. Another basic ingredient is the magnification pattern. In
our problem (magnification near to a straight fold caustic), the magnification 
law is simple and well-known. Taking a cartesian coordinate frame in which the 
caustic line is defined by the $y$-axis, the magnification of a piece of the 
disk at ($x$,$y$) will be (e.g., Schneider \& Weiss 1987)  
\begin{equation} 
A(x) = A_0 + a_C H(x)/\sqrt{x}  ,
\end{equation}
where $H(x)$ = 1 at $x >$ 0 and $H(x)$ = 0 at $x \leq$ 0.
In the new coordinate frame, the magnification law is only function of the
coordinate $x$, and we can easily relate this cartesian coordinate to the
corresponding ones ($r$,$\psi$) in the natural coordinate frame of the disk. If
the centre of the source is placed at $x = x_c$, we found a relationship
between $x$ and ($r$,$\psi$): 
\begin{equation} 
x = x_c + r \sin \psi \cos \alpha \sin \beta - r \cos \psi \cos \beta .
\end{equation}
To derive Eq. (4) it was considered that the axis of the disk is inclined by an
angle $\alpha$ in the plane of the observer's sky, and the node line and the
$x$-axis are separated by an angle $\beta$. The elemental flux from the surface
element $r dr d\psi$ can be now estimated in a direct way. In the absence of
the lens, $d\Omega_* = D_s^{-2} r dr d\psi \cos \alpha$ is the solid angle
subtended in the observer's sky by the piece of the disk, where $D_s$ is the 
angular diameter distance to the source. The true solid angle will be $d\Omega
= A[x(r,\psi;x_c)] d\Omega_*$, and consequently, the elemental flux has an
expression $dF_{\nu}(r,\psi;x_c) = D_s^{-2} I_{\nu}(r) A[x(r,\psi;x_c)] r dr 
d\psi \cos \alpha$. Moreover, the integral of 
$dF_{\nu}(r,\psi;x_c)$ over the whole disk gives the total monochromatic flux 
$F_{\nu}(x_c)$, i.e., 
\begin{equation} 
F_{\nu}(x_c) = K_{\nu} \int_{r_{in}}^{r_{out}} 
\frac{r dr}{\exp [h\nu/kT(r)] - 1} 
\int_{0}^{2\pi} \{1 + \frac{a_C}{A_0}
\frac{H[x(r,\psi;x_c)]}{\sqrt{x(r,\psi;x_c)}}\} d\psi   ,
\end{equation}
where $K_{\nu} = (2h\nu^3/c^2D_s^2)A_0 \epsilon_{\nu} \cos \alpha$, and
$r_{out}$ is the outer edge of the disk, which ($r_{out}$) could be of the
order of 100--300 inner radii ($r_{in}$). 

In order to derive the microlensing light curve from Eq. (5), we firstly
calculated the integral over $\psi$. The angular integration led to
\begin{equation} 
\int_{0}^{2\pi} [1 + \frac{a_C}{A_0} \frac{H(x)}{\sqrt{x}}] d\psi =
2\pi \left[ 1 + \frac{a_C}{\pi A_0 \sqrt{rf}} G(q) \right]  ,
\end{equation}
where $f = (\cos ^2 \beta + \cos ^2 \alpha \sin ^2 \beta)^{1/2}$, $q = x_c/rf$,
and 
\begin{equation} 
G(q) = \int_{-1}^{+1} \frac{H(q - y) dy}{\sqrt{q - y} \sqrt{1 - y^2}}   .
\end{equation}
Inserting all these results into Eq. (5), one obtains 
\begin{equation} 
F_{\nu}(x_c) = 2\pi K_{\nu} \int_{r_{in}}^{r_{out}} 
[1 + \frac{a_C}{\pi A_0 \sqrt{rf}} G(x_c/rf)] 
\frac{r dr}{\exp [h\nu/kT(r)] - 1}    .
\end{equation}
In a second step, we used the new variable $\xi = r/r_{in}$ as well as the
trajectory of the centre of the source $x_c(t) = V_{\perp}(t - t_0)$. As usual,
$V_{\perp}$ is the quasar velocity perpendicular to the caustic line, and $t_0$
is the time of caustic crossing by the source centre. The final light curve has
an expression
\begin{equation} 
F_{\nu}(t) = A_{\nu} \int_{1}^{\xi_{max}} 
\{1 + \frac{B}{\sqrt{\xi}} G[C(t - t_0)/\xi]\} 
\frac{\xi d\xi}{\exp \left[0.054 \left(\frac{\nu}{10^{14} \hspace{0.07in} 
\rm{Hz}}\right) D \xi^{3/4} (1 - 1/\sqrt{\xi})^{-1/4}\right] - 1}     ,
\end{equation}
where
\begin{equation} 
A_{\nu} = \frac{2\pi r_{in}^2}{D_s^2} \left(\frac{2h\nu^3}{c^2}\right) A_0 
\epsilon_{\nu} \cos \alpha  ,
\end{equation}
\begin{equation} 
B = \frac{a_C}{\pi A_0 (r_{in}f)^{1/2}}  ,
\end{equation}
\begin{equation} 
C = \frac{V_{\perp}}{r_{in}f}  ,
\end{equation}
\begin{equation} 
D = \left(\frac{M}{10^8 M_{\odot}}\right)^{1/2} \left(\frac{dM/dt}{10^{26}
\hspace{0.07in} \rm{g} \hspace{0.07in} \rm{s}^{-1}}\right)^{-1/4}  .
\end{equation}
In Eq. (9), $\xi_{max}$ is the ratio between the outer radius and the inner
radius, i.e., $\xi_{max} = r_{out}/r_{in}$. Moreover, in order to reduce the
computing time, from some MATHEMATICA packages, we inferred an analytical
approximation to the function $G(q)$ (see Appendix A). The exact function is 
very well traced by the approximated one, with a typical accuracy of about 
0.01\%. We note that the time evolution of the theoretical HME depends on the 
frequency, two chromatic amplitudes ($A_{\nu}$, $B_{\nu} = B A_{\nu}$), and 
four achromatic parameters ($\xi_{max}$, $C$, $D$, and $t_0$).

In practice one must compare the theoretical law (9) with the brightness record
of a QSO image during a HME. In the comparison, the frequency $\nu$ will be the
central frequency of the optical band in which the observations were made, and
as mentioned here above, $\xi_{max}$ = 100 ($r_{out}$ = 300 Schwarzschild
radii) or $\xi_{max}$ = 300 ($r_{out} \sim$ 1000 Schwarzschild radii) are
plausible choices of the edge ratio. Therefore, at a given $\xi_{max}$ (e.g.,
100), the goal is to estimate the values of the parameters $A_{\nu}$, $B_{\nu}$,
$C$, $D$, and $t_0$ by fitting the observed HME. Once the parameters $C$ and $D$
have been measured, from Eqs. (12-13) we make the relations: 
\begin{equation} 
\mu_{BH} = 10^{-4} \hspace{0.07in} V_{\perp}({\rm km} \hspace{0.07in} 
{\rm s}^{-1}) / [fC \hspace{0.07in} ({\rm day}^{-1})] ,
\end{equation}
\begin{equation} 
\mu_{AC} = \mu_{BH}^2/D^4  ,
\end{equation}
with $\mu_{BH} = M/10^8 M_{\odot}$ and $\mu_{AC} = (dM/dt)/10^{26} 
\hspace{0.07in} \rm{g} \hspace{0.07in} \rm{s}^{-1}$. Although the factor
related to the orientation of the source ($f$) cannot be directly obtained from
the fit, its estimation is relatively simple. We simulated 10$^4$ pairs of 
values ($\alpha$,$\beta$) covering the whole range of possible orientations, 
and calculated the corresponding $f$ values. From the simulations, we got and 
drew (Fig. 1) the probability distribution $P(f)$. In Fig. 1 it is apparent 
that the bulk of the probability is associated with high values of $f$, and the
maximum of the distribution is reached when $f \rightarrow$ 1. Only very 
particular orientations (nearly edge-on disks with a node line almost parallel 
to the caustic line) lead to small values of the factor $f$. From the 
probability distribution we inferred that $<f>$ = 0.84 and $f = 
0.84^{+0.16}_{-0.02}$ at 1$\sigma$ confidence level (68.9 percent of the 
simulations; see the shading region in Fig. 1). Finally, in order to measure
the central mass and the flow of matter, it is only required dynamical
information. When that information is available, through Eqs. (14-15) one can
estimate both physical quantities. 
%------------------------------------------FIGURE1
\begin{figure}[hbtp]
 \centering
 \epsfxsize=10.00 cm
 \rotatebox{-90}{\epsffile{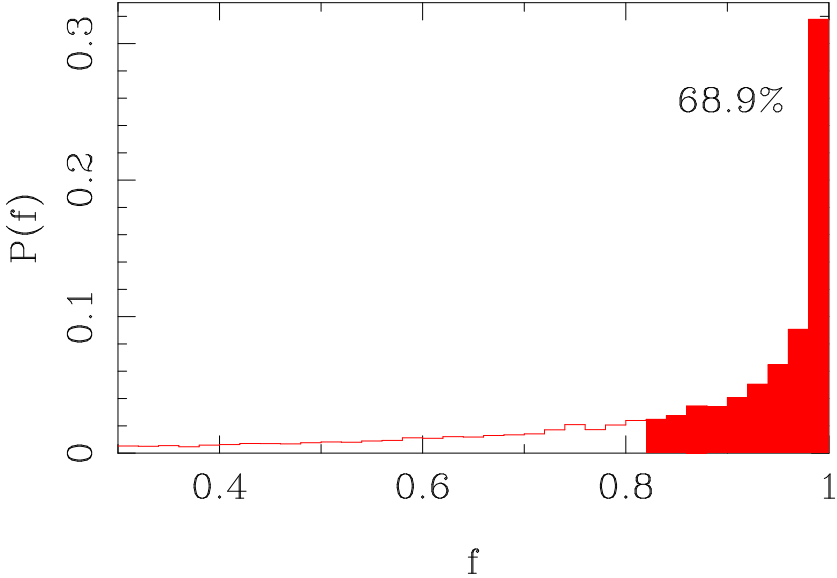}}
 \caption[]{Probability distribution of the factor related to the
orientation of the source. The shading region represents the 1$\sigma$
confidence interval including both the mean value and the most probable
one.}
 \label{f1} 
\end{figure}
%------------------------------------------ 

\section{Physical properties of the accretion disk in QSO 2237+0305}

The GLITP collaboration has recently reported on the peak of a microlensing
high-magnification event detected in both $V$-band and $R$-band brightness
records of Q2237+0305A, which was observed with an excellent sampling rate of 
about three times the OGLE one (Alcalde et al. 2002). The event was discovered 
by the OGLE team (Wo{\' z}niak et al. 2000), who have followed the evolution of
the whole $V$-band HME. In a first paper, from the GLITP $V$-band and $R$-band
peaks, Shalyapin et al. (2002) have analyzed the nature and size of the
$V$-band and $R$-band sources in QSO 2237+0305. To do this task, they used a
family of source models. All the models led to theoretical microlensing curves 
with the same number of free parameters. This approach (comparison between models 
causing four-parametric theoretical curves) is very useful to discuss the 
feasibility of different
scenarios. The authors concluded that only two rough versions of the Newtonian
standard scenario are clearly consistent with the observed optical peaks. Thus
we know that a Newtonian standard accretion disk is in agreement with the GLITP
monitoring of Q2237+0305A, and in principle, it is possible to measure the
central mass and the flow of matter (see Sect. 2).  

\subsection{Measurements of $fC$ and $D$}

In our microlensing experiment, the functional relation presented in Eq. (9) is
investigated by comparing it with the observed value of $F_{\nu}(t)$ at a given
frequency and for various epochs $t_1,t_2,...,t_N$. A value of $\xi_{max}$ is
also assumed. The goal is to find the parameters $A_{\nu}$, $B_{\nu}$, $C$, $D$, 
and $t_0$ of the theoretical microlensing curve which best describe the
observational data. To make the fit, two different chi-square minimizations
have been carried out. The most elementary procedure is to form a grid of
points in the free parameters and evaluate the $\chi^{2}$ function at each of
these points. The point with the smallest value is then the minimum. Therefore, 
we have studied a reduced grid using some MATHEMATICA packages. The procedure is 
called {\it RGMath} and it is based on the ideas discussed in subsection 2.3 of 
Shalyapin et al. (2002). Basically, as we like to fit a theoretical curve that 
is linear in two parameters ($A_{\nu}$ and $B_{\nu}$), and non-linear 
in $C$, $D$, and $t_0$, the equations of minimization 
$\partial\chi^2/\partial A_{\nu} = \partial\chi^2/\partial B_{\nu} = 0$ lead to 
analytic relations: $A_{\nu} = A_{\nu}(C,D,t_0)$, $B_{\nu} = B_{\nu}(C,D,t_0)$. 
In this scheme, the number of effective parameters is reduced to three, and one 
can work with a 3D grid instead of a 5D one. As the procedure is quite time
consuming, we initially formed a 3D grid with moderate resolution, i.e., the 
size of the grid step was reasonable, but relatively large. So, from {\it RGMath}
at moderate resolution, we inferred an initial best solution and some 
uncertainties associated with it. Another procedure is called {\it DSFor} and it 
is related to a version of the simplex method. We made a FORTRAN program to apply
a downhill simplex methodology (Nelder \& Mead 1965) to our problem. The
simplexes will be the simplest geometrical figures in 5 dimensions having 6
vertices (points). The technique begins by choosing a point (vertice) in the 5D
parameter space. Then the program generates an initial simplex, and after,
repeated calculations are made while varying the vertices of the initial
simplex in some way, until a local minimum is reached. The $\chi^{2}$ function
may have different local minima around different points tested from the method,
and consequently, to find the global minimum $\chi^{2}(min)$ we must make a 5D
grid, and apply the downhill simplex technique to all the points in the grid.
In the first stage with this alternative task (search of minima in the 5D 
parameter space), the resolution was not very good. Once the initial best solution 
$~\vec{v^*}$ = ($A_{\nu}^*$, $B_{\nu}^*$, $C^*$, $D^*$, $t_0^*$) is known (via
{\it RGMath} or {\it DSFor}), it is possible to refine it and estimate accurate 
errors for the relevant parameters $C$ and $D$. If the vector $~\vec{v}$ = 
($A_{\nu}$, $B_{\nu}$, $C$, $D$, $t_0$) of parameter values is perturbed away from 
$~\vec{v^*}$, then $\chi^{2}$ increases. So we can draw high-resolution "parabolic"
laws $\chi^{2}_{C}(min)$ and $\chi^{2}_{D}(min)$, where, for example, 
$\chi^{2}_{C}(min)$ represents the minima of the $\chi^{2}$ function for values 
of $C$ around $C^*$. The $C$ range within which $\chi^{2}_{C}(min) - \chi^{2}(min)
\leq$ 1 defines the 1$\sigma$ confidence interval in the estimation of $C$, and
the interval $\chi^{2}_{D}(min) - \chi^{2}(min) \leq$ 1 is related to the 
1$\sigma$ confidence interval in the estimation of $D$. The $k\sigma$ confidence 
interval in the generic parameter $p$ corresponds to the region bounded by 
$\chi^{2}_{p}(min) - \chi^{2}(min)$ = $k^2$ ($k$ = 1,2,...). 

%------------------------------------------TABLE1
\begin{table}
\centering
\begin{tabular}{ccccccc}
\hline\noalign{\smallskip}
Procedure & $\xi_{max}$ & $A_{\nu}$ ($\mu$Jy) & $B_{\nu}$ ($\mu$Jy) &
$fC$ (day$^{-1}$) & $D$ & $t_0$ (JD--2450000) \\
\noalign{\smallskip}\hline\noalign{\smallskip}
{\it RGMath} & 100 & 2.78 & 1.37 & 
$0.23^{+0.13}_{-0.06}$ & $0.59^{+0.30}_{-0.23}$ & 1481.4 \\
{\it DSFor} & 100 &  2.96 & 1.35 & 
$0.23^{+0.14}_{-0.08}$ & $0.60^{+0.35}_{-0.26}$ & 1481.9 \\
\noalign{\smallskip}\hline
\end{tabular}
\caption{Parameters of the GLITP $V$-band microlensing peak.\label{tbl-1}}
\end{table}
%------------------------------------------
%------------------------------------------FIGURE2
\begin{figure}[hbtp]
 \centering
 \epsfxsize=10.00 cm
 \rotatebox{-90}{\epsffile{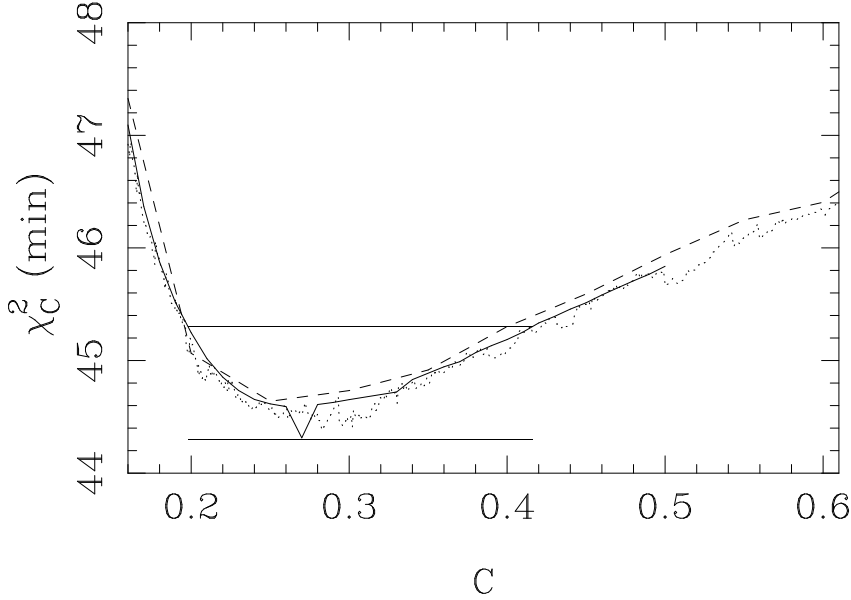}}
 \caption[]{Behaviour of $\chi^{2}_{C}(min)$ derived from the GLITP
$V$-band light curve for Q2237+0305A. The results with {\it RGMath} are 
shown by a dashed line (moderate resolution) and a solid line (high resolution).
The dotted line was obtained with the {\it DSFor} task (high resolution).
It is also depicted the 1$\sigma$ box corresponding to the {\it RGMath}  
procedure (high resolution).}
 \label{f2} 
\end{figure}
%------------------------------------------ 
The high-resolution fit to the GLITP $V$-band light curve for Q2237+0305A appear 
in Table 1. The central frequency is of $\nu_V$ = 5.52 10$^{14}$ Hz, and we 
adopted $\xi_{max}$ = 100. We note that the uncertainties presented in Table 1 
are 1$\sigma$ intervals. Moreover, instead of direct measurements of the parameter 
$C$, 1$\sigma$ confidence intervals for the relevant factor $fC$ (see Eq. 14) are 
quoted in Table 1. During an advanced stage of the project, we also obtained rough 
estimates of the errors in $A_{\nu}$ and $B_{\nu}$. From {\it DSFor} we inferred 
$A_{\nu} \sim 3^{+6}_{-1}$ $\mu$Jy and $B_{\nu} \sim 1.5^{+2.6}_{-0.6}$ $\mu$Jy. 
In Fig. 2 we show the details to obtain the best values of $C$ and the standard 
errors. The figure presents three $\chi^{2}_{C}(min)$ trends, which were derived 
from {\it RGMath} at moderate resolution (dashed line), {\it RGMath} at high 
resolution (solid line), and {\it DSFor} at high resolution (dotted line). The 
1$\sigma$ box corresponding to the {\it RGMath} (high resolution) procedure is 
plotted with a thin line. Looking both Table 1 and Fig. 2, two important conclusions 
appear: (1) the fit is stable against a change in the fitting method ({\it RGMath} 
or {\it DSFor}), and (2) a change from moderate to high resolution does not strongly 
perturb the parabolic law. Taking into account these last conclusions, we used the 
{\it RGMath} task (moderate resolution) to test the influence of the $\xi_{max}$ 
value in the parameter estimation. We concluded that a change in $\xi_{max}$ (from 
100 to 300) does not modify the estimates of $fC$ and $D$. In principle, a fit to 
the GLITP $R$-band light curve for Q2237+0305A could be useful to improve the 
parameter estimation. However, the GLITP $R$-band microlensing peak does not permit 
to infer the parameters $C$ and $D$ with errors similar to the previous ones (from 
the $V$-band peak). The ratio between the 1$\sigma$ $R$-band interval and the 
1$\sigma$ $V$-band interval is $>$ 2.4 for $C$ and $\approx$ 1.5 for $D$. Due to 
these relatively large errors from the $R$-band brightness record, we only
considered the estimates in Table 1. More properly, we took $fC = 
0.23^{+0.13}_{-0.06}$ day$^{-1}$ and $D = 0.59^{+0.30}_{-0.23}$ ({\it RGMath}
solution) to determine the possible values of $\mu_{BH}$ and $\mu_{AC}$. We
remark that the accretion-disk model is consistent with the $V$-band and $R$-band 
light curves. The best-fit reduced $\chi^{2}$ is very close to one in both optical 
filters, and we can see the good agreement between best-fits and observational 
trends in Fig. 3. In the top panel, the observed $V$-band fluxes are compared with
the best-fit from {\it RGMath} at high resolution (solid line), while in the
bottom panel, the $R$-band record is compared with the best-fit from {\it RGMath} 
at moderate resolution (dashed line). The fit in the $R$ band led to best values
of $C$ = 0.375 day$^{-1}$ and $D$ = 0.45, which are totally consistent with the
{\it blue} parameters given in Table 1. However, the $R$-band data are relatively 
noisy in comparison to the $V$-band ones, and this fact does not permit a parameter
estimation in the {\it red} band with uncertainties similar to the {\it blue} 
errors.  

%------------------------------------------FIGURE3
\begin{figure}[hbtp]
 \centering
 \epsfxsize=10.00 cm
 \rotatebox{0}{\epsffile{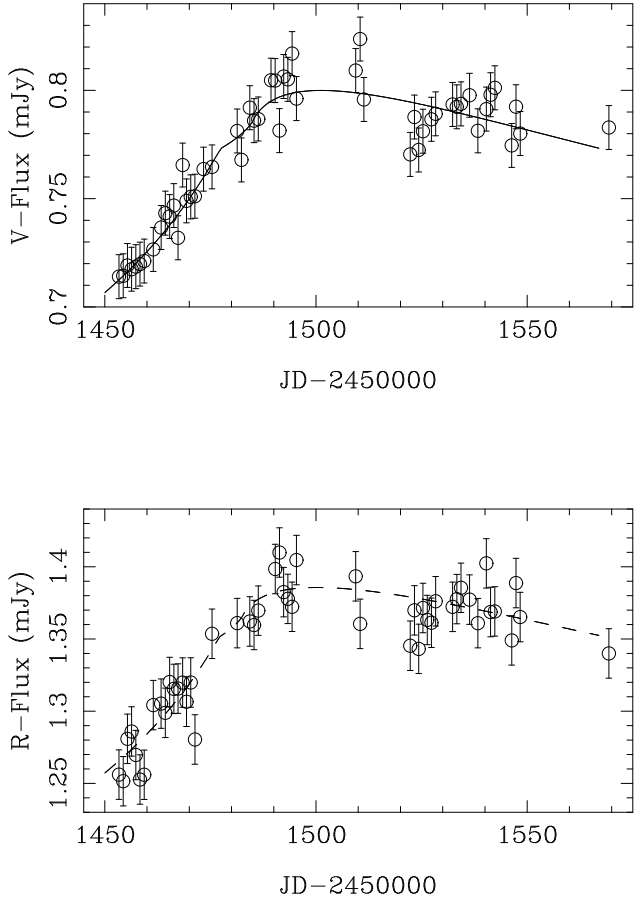}}
 \caption[]{Observed fluxes and best-fits. In the top panel ($V$-band), it is 
showed the best-fit from {\it RGMath} at high resolution (solid line). In the
bottom panel ($R$-band), it appears the best-fit from {\it RGMath} at moderate 
resolution (dashed line).}
 \label{f3} 
\end{figure}
%------------------------------------------ 

\subsection{Adoption of a dynamical range}

Kayser et al. (1986) showed that the effective transverse velocity of the
source consists of three terms related to the transverse peculiar velocity of
the source quasar ($~\vec{v_s}$), the transverse peculiar velocity of the 
deflector ($~\vec{v_d}$), and the transverse peculiar motion of the observer 
($~\vec{v_o}$). The word "transverse" denotes "perpendicular to the line of
sight", and the effective transverse velocity is given by 
\begin{equation} 
{\bf V} = \frac{~\vec{v_s}}{1 + z_s} - \frac{~\vec{v_d}}{1 + z_d}\frac{D_s}{D_d} 
+ \frac{~\vec{v_o}}{1 + z_d}\frac{D_{ds}}{D_d} ,
\end{equation}                               
where $z_d$ is the redshift of the deflector (lens galaxy), $D_d$ is the
angular diameter distance to the deflector, and $D_{ds}$ is the angular
diameter distance between the deflector and the source. In our case (QSO
2237+0305), the deflector is a nearby spiral galaxy ($z_d$ = 0.039 $<<$ 1) and
the source is a far quasar ($z_s$ = 1.695). Due to these facts, one has that
$(1+z_s)/(1+z_d)$ = 2.6 and $D_{ds}/D_d \sim D_s/D_d \sim$ 10, and the 
particular configuration permits to neglect the term caused by the peculiar
motion of the source. With respect to the motion of the observer, from the
dipole observed in the cosmic microwave background (CMB), it is derived a
peculiar velocity of the Sun: $v_{obs}$ = 369 km s$^{-1}$ in the direction
(J2000) $\alpha$ = 11$^{\rm h}$11$^{\rm m}$57$^{\rm s}$, $\delta$ = -- 
7.$^{\circ}$22 (Lineweaver et al. 1996). This peculiar motion is almost
perpendicular to the position for the system, being $|~\vec{v_o}| \approx$ 50 km
s$^{-1}$. Because of the small observer's motion (as compared with the expected
transverse peculiar velocity of the nearby spiral), we can also neglect the
last term in Eq. (16). Therefore, for QSO 2237+0305, the effective transverse
velocity takes the very simple form
\begin{equation} 
~\vec{V} = - \frac{~\vec{v_d}}{1 + z_d}\frac{D_s}{D_d}  .
\end{equation}                               
We are indeed interested in the effective quasar velocity perpendicular to the
microcaustic of interest. Thus, if $~\vec{i}$ is the unit vector in the
direction perpendicular to the caustic line, then
\begin{equation} 
V_{\perp} = ~\vec{V.i} = \frac{v_d}{1 + z_d}\frac{D_s}{D_d}  ,
\end{equation}          
where $v_d = |~\vec{v_d.i}|$ is the amplitude of the projected peculiar
motion of the deflector. 

Only one direct measurement of $V = |~\vec{V}|$ is currently available. The 
measurement is an upper limit on $V$, and consequently on $V_{\perp}$, which was
inferred from the analysis of the light curves of the system (Wyithe, Webster \& 
Turner 1999). So we have decided to adopt a velocity range based on the relation 
(18) and the relevant observational data. Firstly, one must find plausible values 
of $v_d$, and with this aim, we analyzed a catalog of galaxy data that is included 
in the electronic archives of the CDS (Centre de Donn\'ees astronomiques de
Strasbourg). The catalog is a part of the Mark III full catalog of redshifts and 
distances (Willick et al. 1997), and it contains redshifts 
in the CMB frame and both forward and inverse Tully-Fisher (TF) distances of 1355
spiral galaxies (Mathewson, Ford \& Buchhorn 1992). As it is well-known, the 
differences $z_{CMB} - r_{TF}$ are the projected peculiar motions of the spirals, 
i.e., $~\vec{v_g.n_g}$, where $~\vec{n_g}$ are the unit vectors pointing
towards the galaxies. In Fig. 4 we can see two very similar histograms showing 
two distributions of $v_g = |~\vec{v_g.n_g}|$. The darkish line represents 
the results from the inverse TF distances ($r_{ITF}$), while the other line 
traces the behaviour from the forward TF distances ($r_{FTF}$). Inhomogeneous 
Malmquist bias-corrected distances $r_{FTF}$ are quoted in the catalog, which
were computed using the density field obtained through the IRAS 1.2 Jy redshift
survey. From the distributions in Fig. 4, we concluded that the mean value is of
$<v_g>$ = 663 km s$^{-1}$. This average will be within the interval of allowed
values of $v_d$. As a lower limit we took 100 km s$^{-1}$, i.e., $v_d \geq$ 100 
km s$^{-1}$, in agreement with the observed distributions [$P$($v_g \leq$ 50 km 
s$^{-1}$) $\leq$ 6\%] and discarding a conspiracy of nature (a situation in 
which both $v_o = |~\vec{v_o.i}|$ 
and $v_d$ have a value $\leq$ 50 km s$^{-1}$). On the other hand, the 
distributions $P(v_g)$ were obtained with a sample including field spirals and
cluster spirals, and it is evident that amplitudes of the projected peculiar 
velocities in excess of 1000 km s$^{-1}$ cannot be rejected. However, the lens 
galaxy is not placed inside a rich cluster or close to a group of clusters,
and thus, one may assume an upper limit of 1000 km s$^{-1}$, i.e., 
$v_d \leq$ 1000 km s$^{-1}$. To derive a reasonable range for $V_{\perp}$, the 
second basic ingredient is the factor $D_s/(1+z_d)D_d$. This factor depends on 
the redshifts ($z_d$ and $z_s$) and the matter/energy content of the universe 
(cosmological parameters). In this paper we consider the two standard flat
cosmologies: (1) $\Omega_0$ = 1, $\lambda_0$ = 0, and (2) $\Omega_0$ = 0.3, 
$\lambda_0$ = 0.7. As usual, $\Omega_0$ is the present density parameter and
$\lambda_0$ is the present reduced cosmological constant. So, each plausible 
value of $v_d$ leads to two solutions for $V_{\perp}$. Taking into account 
the whole range 100 $\leq$ $v_d$ (km s$^{-1}$) $\leq$ 1000, the characteristic 
redshifts of the system, and the two standard cosmologies, it is found that 
765 $\leq$ $V_{\perp}$ (km s$^{-1}$) $\leq$ 10548.
%------------------------------------------FIGURE4
\begin{figure}[hbtp]
 \centering
 \epsfxsize=10.00 cm
 \rotatebox{-90}{\epsffile{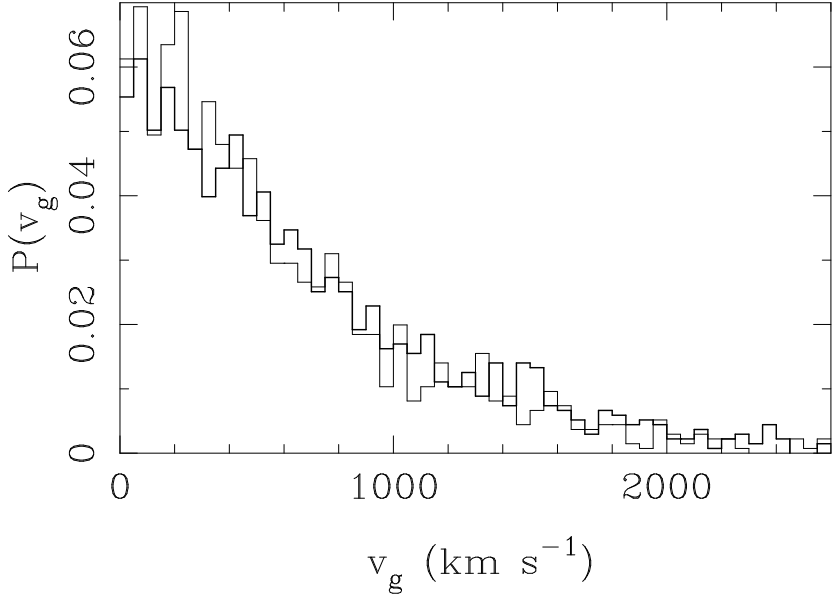}}
 \caption[]{Probability distributions of the amplitudes of the
projected peculiar motions of 1355 nearby spiral galaxies (see Mathewson, 
Ford \& Buchhorn 1992). The darkish line corresponds to the results from
the inverse Tully-Fisher distances, and the other line represents the results
from the forward Tully-Fisher distances, which were corrected for inhomogeneous 
Malmquist bias.}
 \label{f4} 
\end{figure}
%------------------------------------------ 

\subsection{Results}

Our results (1$\sigma$ measurements of $M$ and $dM/dt$ for two 
cosmologies and different values of $v_d$) are presented in Table 2. We tested
four representative values of $v_d$ (see subsection 3.2). The black hole mass is 
reasonably well constrained. For a universe with a zero cosmological
constant and a very small projected motion of the deflector,  
the microlensing data suggest the existence of a central black hole having a
mass of $\approx$ 3.3 10$^7$ $M_{\odot}$. The most massive black hole is 
inferred for a universe with a nonzero cosmological constant ($\lambda_0$ = 0.7)
and a very large projected velocity of the deflector. In that case, $M 
\approx$ 4.6 10$^8$ $M_{\odot}$. Apart from these "individual" estimates, 
the main result is the global range for $M$. We concluded that the dark
mass in the heart of the quasar should be larger than 10$^7$ $M_{\odot}$
and smaller than 6 10$^8$ $M_{\odot}$.

With regard to the mass accretion rate, the situation is quite intricate. 
Firstly, we cannot obtain individual lower limits, since all the 1$\sigma$
estimates are consistent with $dM/dt$ = 0. Therefore, only typical values
and upper limits are showed in Table 2. The fact that the 1$\sigma$ error 
bars on the accretion rate turn out to be compatible with zero is related to the 
(dis)abilities of the new technique. We fitted the parameters $C$ and $D$ (together 
with $A_{\nu}$, $B_{\nu}$, and $t_0$), and used Eqs. (14-15) to estimate the 
physical properties of the central engine: $M$ and $dM/dt$. Considering the 
results in Table 1, $fC$ is determined with a relative error of about 40 per cent,
while the relative error in $D$ is of about 45 per cent. If we take a given value 
of the cosmological parameters and $v_d$, then the relative error in $M$ will be 
of about 40 per cent, i.e., similar to the relative error in $fC$ (see Eq. 14 
and the results on $M$ in Table 2). However, unfortunately, Eq. (15) indicates us 
that $dM/dt$ depends on both $(fC)^2$ and $D^4$. Due to this fact, the accuracy 
in the determination of $dM/dt$ is very poor, or in other words, the relative 
error in $dM/dt$ is greater than 100 per cent. 

Secondly, a realistic scenario should incorporate the electron-scattering 
opacity, which could play an important role at inner regions of the accretion disk 
(Shakura \& Sunyaev 1973). The local radiation energy flux is determined by the 
gravitational energy release $Q_{grav} = (3/8\pi) (GM/r^3) (dM/dt)[1 - 
(r_{in}/r)^{1/2}]$. In the outer regions, where the free-free 
processes give the main contribution to the opacity, it is formed a Planck 
spectrum. The radiation energy flux must be $Q_{rad} = \sigma T_s^4$, and from 
$Q_{rad} = Q_{grav}$ we obtain the standard thermal law. In the inner regions of 
the disk, the electron-photon scattering presumably plays a role in the opacity, 
so these regions radiate less efficiently than a blackbody: $I_{\nu_s}[T_s(r)] = 
\varepsilon_{in}[\nu_s,T_s(r)] B_{\nu_s}[T_s(r)]$ with 
$\varepsilon_{in}[\nu_s,T_s(r)] <$ 1. If one defines a constant factor 
$\varepsilon_{in} <$ 1 to be the emissivity relative to a blackbody, then the 
corresponding radiation energy flux will be $Q_{rad} = \varepsilon_{in} \sigma 
T_s^4$. At small radii the real disk could be hotter than the standard disk without 
electron-photon scattering, i.e., $T_s(r) = [(3/8\pi) (GM/\sigma r^3) 
\varepsilon_{in}^{-1} (dM/dt) (1 - \sqrt{r_{in}/r})]^{1/4}$. Therefore we can
consider two simple pictures: a global blackbody spectrum ($\varepsilon$ = 1
along all the disk), i.e., the standard one, and a global greybody spectrum 
($\varepsilon <$ 1 along all 
the disk). This last simplification was suggested by Rauch \& Blandford (1991),
who introduced the greybody spectra in microlensing studies. When a generalized 
model is considered ($\varepsilon \leq$ 1), we must reinterpret the results on 
$dM/dt$ as measurements of $\varepsilon^{-1}(dM/dt)$. In this way, the typical 
values of $dM/dt$ vary in the interval $(1-300)\varepsilon$ $M_{\odot}$ yr$^{-1}$. 
For $\varepsilon$ = 1, we derive a typical range of 1 -- 300 $M_{\odot}$ yr$^{-1}$, 
while for $\varepsilon$ = 0.1, the typical mass accretion rates are 0.1 -- 30 
$M_{\odot}$ yr$^{-1}$.
%------------------------------------------TABLE2
\begin{table}
\centering
\begin{tabular}{cccc}
\hline\noalign{\smallskip}
Cosmology & $v_d$ (km s$^{-1}$) & $M$ (10$^8$ $M_{\odot}$) & 
$dM/dt$ ($M_{\odot}$ yr$^{-1}$) \\
\noalign{\smallskip}\hline\noalign{\smallskip}
$\Omega_0$ = 1, $\lambda_0$ = 0 & 100 & $0.33^{+0.09}_{-0.19}$ & 1.5 ($<$ 3.9) \\
                                & 300 & $1.00^{+0.26}_{-0.56}$ & 13.2 ($<$ 34.8) \\
                                & 663 & $2.21^{+0.58}_{-1.25}$ & 64.3 ($<$ 170.0) \\
				& 1000 & $3.33^{+0.87}_{-1.88}$ & 146.2 ($<$ 386.7) \\
$\Omega_0$ = 0.3, $\lambda_0$ = 0.7 & 100 & $0.46^{+0.12}_{-0.26}$ & 2.8 ($<$ 7.3) \\
                                & 300 & $1.38^{+0.36}_{-0.78}$ & 25.0 ($<$ 66.1) \\
                                & 663 & $3.04^{+0.79}_{-1.72}$ & 122.1 ($<$ 322.8) \\
				& 1000 & $4.59^{+1.20}_{-2.59}$ & 277.7 ($<$ 734.4) \\
\noalign{\smallskip}\hline
\end{tabular}
\caption{Black hole mass and mass accretion rate in QSO 2237+0305.\label{tbl-2}}
\end{table}
%------------------------------------------

\section{Conclusions and discussion}

We showed the time evolution of a microlensing HME caused by a generic 
Newtonian geometrically-thin and optically-thick standard accretion disk 
crossing a generic caustic straight line. Given a lensed QSO, in order to
estimate the mass of its central black hole and the accretion rate, we can
thus compare the theoretical law with a gravitational microlensing HME observed
in some image of the far source. The technique was applied to the gravitational 
mirage QSO 2237+0305, whose brightest image experienced an important change in
flux during the 1999-2000 seasons (Wo{\'z}niak et al. 2000). The peak of the
dramatic variation in Q2237+0305A (Alcalde et al. 2002) led to very interesting 
information on the black hole in the core of the source and an interval of 
typical values for the mass accretion rate. The main source of uncertainty 
comes from our lack of knowledge about the peculiar motion of the lens galaxy
in the direction perpendicular to the caustic line ($v_d$). To accurately 
determine the typical flow of matter, it is also necessary to know the physical
processes involved in the emissivity of the disk. Although the standard 
scenario does not incorporate the electron-photon scattering, in general, it 
could strongly distort the standard Planck spectrum at the innermost layers of 
the disk (e.g., Shakura \& Sunyaev 1973; Malkan 1983). 

From 1$\sigma$ estimates of the relevant parameters in the theoretical 
microlensing curve, we inferred 1$\sigma$ measurements of the black hole mass
and the accretion rate for a set of $v_d$ values, which represents the whole
range of reasonable choices (see a detailed discussion in subsection 3.2). 
Taking into account all the individual estimates in Table 2, one obtains that
QSO 2237+0305 contains a massive black hole: 10$^7$ $M_{\odot}$ $< M <$ 6 
10$^8$ $M_{\odot}$. The information about the mass accretion rate is very much
poor. The main result consists of a typical interval for $dM/dt$. This interval 
is of $(1-300)\varepsilon$ $M_{\odot}$ yr$^{-1}$, where $\varepsilon$ = 1 if 
the free-free processes are dominant (standard source) and $\varepsilon <$ 1 
when the scattering plays a role (e.g., Rauch \& Blandford 1991). In
spite of the fact that the range of typical determinations of $dM/dt$ is not
excessively broad, all the individual 1$\sigma$ measurements are in agreement
with no accretion but a flow of matter of almost $1000 \varepsilon$ $M_{\odot}$ 
yr$^{-1}$ cannot be ruled out. On the other hand, one direct constraint on 
$v_d$ was recently reported by Wyithe, Webster \& Turner (1999). This group
(Wyithe, Webster, Turner and other colleagues) is involved in a project to
interpret the QSO 2237+0305 microlensing light-curves, and as a part of the
program, they derived an upper limit of $v_t <$ 500 km s$^{-1}$, where $v_t$   
is the galactic transverse velocity ($\Omega_0$ = 1, $\lambda_0$ = 0). 
Therefore, considering that $v_d <$ 500 km s$^{-1}$ and $v_d \geq$ 100 
km s$^{-1}$, we find a central value of $v_d$ = 300 km s$^{-1}$. For a flat
universe with a zero cosmological constant and a projected motion of $v_d$ = 
300 km s$^{-1}$, the typical source parameters are: $M$ = 10$^8$ $M_{\odot}$
and $dM/dt \approx$ $10\varepsilon$ $M_{\odot}$ yr$^{-1}$.

Several modeling aspects can be improved, anyway. For example, one must
include the effects of general relativity. One may also incorporate a new
theoretical law to the gallery of models: the time evolution of a microlensing
event generated by an accretion disk passing close to a cusp caustic. Efforts
in both directions are now in progress, and we will consider these improvements
in future studies of new QSO 2237+0305 microlensing peaks. Fortunately, using
the present model and monitoring the events that occur in the components of the
system, the future prospects are very promising. Thus, from 10 microlensing 
peaks monitored in ten different optical bands (blue bands are required to avoid 
the possible contamination by other sources of light as the extrapolated 
IR power law spectrum, and therefore, the possible perturbation of the chromatic 
amplitude $A_{\nu}$ which is defined in Eq. (10) for a blackbody emission; see, 
for example, Malkan 1983), we must be able to noticeably
reduce the current uncertainty in the microlensing parameter $D$: from about
0.3 (see Table 1) to 0.03. In a similar way, if each peak is recorded at 10
frequencies, the error in each $C$ value will be lowered in a factor of about
3, i.e., from a mean uncertainty of $\approx$ 0.1 day$^{-1}$ to $\approx$ 0.03 
day$^{-1}$. As a result of new detailed monitoring programs, accuracies of 5-10\%
in the microlensing parameters $C$ and $D$ can be easily achieved in the next 
years. If a global dataset is made for each microlensing peak (tracing the time
evolution of the spectrum), and it is fitted to a model with six free parameters:
$A$, $B$, $C$, $D$, $t_0$, and $\gamma$ ($A_{\nu}$ = $A \nu^{3+\gamma}$, $B_{\nu}$ 
= $B \nu^{3+\gamma}$), the accuracy could be even better than a few percent. 
However, as it was previously remarked, we need a robust estimation of
the lens galaxy motion to accurately measure the black hole mass. This "dynamical
problem" is the only pitfall to obtain a robust measurement of the amount of
dark mass in the centre of QSO 2237+0305. From another point of view, if an
independent estimate of $M$ were available, we would deduce the effective quasar
velocity $V_{\perp}$ involved in the different microlensing events as well as 
the value of $\varepsilon^{-1}(dM/dt)$. Through a reasonable collection of
$V_{\perp}$ data, one may derive the effective transverse velocity of the 
quasar ($V$), which is a basic piece in some microlensing analyses. Up to now
we focussed on the parameters ($C$, $D$) and the properties of the source,
however, the parameters ($A_{\nu}$, $B_{\nu}$) are also relevant for other studies.
For example, if it is available a multiband monitoring of a microlensing peak,
we may estimate the ratios $A_{\nu(i)}/A_{\nu(j)}$. These ratios will be 
directly related to the extinction ratios $\epsilon_{\nu(i)}/\epsilon_{\nu(j)}$,
and consequently, to the global (host galaxy + lens galaxy + Milky Way) 
extinction law.

Finally, we wish to put into perspective the new technique to determine the
physical parameters of a far quasar and the measurements for QSO 2237+0305.
Evidence for central black holes in local galaxies comes from small rotating
gaseous and stellar disks discovered with the Hubble Space Telescope, VLBI
observations of H$_2$O masers in Keplerian rotation in some galactic nuclei,
and near-infrared measurements of the radial and proper motions of stars in 
the cluster at the centre of the Galaxy (e.g., Ford et al. 1998). Analyzing 
these gravitational signatures of massive black holes, the typical values of
the black hole masses vary between $\sim$ 3 10$^6$ $M_{\odot}$ for the Milky
Way (Genzel et al. 1997; Ghez et al. 1998) and $\sim$ 3 10$^9$ $M_{\odot}$
for M87 (Ford et al. 1996; Macchetto et al. 1997). The central dark mass can
be also measured in far AGNs using two classical methods: analyses of the blue
and ultraviolet regions of the rest-frame spectra (e.g., Malkan 1983; Sun \&
Malkan 1989) and observations of the time delays between the variations of
the continua and the variations of the broad emission lines (e.g., Wandel, 
Peterson \& Malkan 1999). The reverberation technique provides an estimate of 
the size of the broad-line region (BLR) as well as the central mass of a quasar 
at redshift
$z_{QSO}$. If the line-emitting gas is gravitationally bound to the central 
black hole, the virial theorem implies that $M \propto R_{BLR} \sigma^2$, where 
$\sigma$ is the FWHM of the emission profile of the BLR at distance $R_{BLR}$. 
Therefore, measuring the lag $\tau$ between an intrinsic event at the central 
continuum source and the reply (reverberation) at the BLR, one obtains the size 
$R_{BLR}$ = $c\tau/(1 + z_{QSO})$ and the dark mass $M$ (e.g., Vestergaard 2002
and references therein). Apart from the reverberation mapping technique, there
is a straightforward method to get the parameters $M$ and $dM/dt$. A comparison
between the blue-ultraviolet excess flux of a QSO (over the extrapolated 
infrared power-law spectrum) and the theoretical integrated disk spectrum 
permits to deduce the two physical quantities (e.g., Malkan 1983). In a pioneer
work, Shields (1978) found $M$ = 10$^9$ $M_{\odot}$ and $dM/dt$ = 3 
$M_{\odot}$ yr$^{-1}$ for a Newtonian blackbody disk modelling the central
engine in 3C 273. To make the fits, Malkan (1983) used a relativistic standard 
accretion disk. The theoretical model included the effects of general relativity, 
but the inclination and scattering atmosphere were ignored. For a sample of 6 
quasars and assuming nonrotating black holes, the "big blue bumps" can be fitted
by spectra of disks around massive ($\sim$ 10$^8$ $M_{\odot}$) black holes.
The author also concluded that the six high-luminosity quasars are emitting 
energy at approximately their Eddington limits. The effects of disk inclination 
have been more recently considered by Sun \& Malkan (1989), while the effects 
of electron scattering were discussed by Wandel \& Petrosian (1988). In this
paper we applied an alternative methodology to determine ($M$, $dM/dt$): the 
monitoring of a lensed quasar when a microlensing HME occurs (e.g., Yonehara et 
al. 1998; Agol \& Krolik 1999). The method can be seen as a complementary tool
to prove the existence of massive dark objects and accretion disks in the
centre of far QSOs. Taking as reference values $M$ = 10$^8$ $M_{\odot}$
and $dM/dt \approx$ $10\varepsilon$ $M_{\odot}$ yr$^{-1}$ (see here above), 
the estimate of the central dark mass in QSO 2237+0305 is consistent 
with data of other galaxy nuclei. However, the accretion rate is relatively 
large for a blackbody disk ($\varepsilon$ = 1). As $L/L_{Edd} \propto 
(dM/dt)/M$, where $L$ is the total disk luminosity and $L_{Edd}$ is the 
Eddington luminosity, one obtains $L/L_{Edd} \sim$ 3 for $\varepsilon$ = 1, 
$L/L_{Edd} \sim$ 1 for $\varepsilon$ = 0.3, and $L/L_{Edd} \sim$ 0.3 for 
$\varepsilon$ = 0.1. Therefore, to avoid meaningless results ($L > L_{Edd}$), 
the electron-photon scattering must play a role. 

\begin{acknowledgements}
We would like to thank Vyacheslav Shalyapin for his insights on fitting
procedures. We also thank Joachim Wambsganss and the anonymous referee for 
support during the project and useful comments, respectively.
The GLITP observations were made with Nordic Optical Telescope (NOT), which is
operated on the island of La Palma jointly by Denmark, Finland, Iceland, 
Norway, and Sweden, in the Spanish Observatorio del Roque de Los Muchachos of 
the Instituto de Astrof\'{\i}sica de Canarias (IAC). We are grateful to the
technical team of the telescope for valuable collaboration during the 
observational work. This work was supported by the P6/88 project of the IAC, 
Universidad de Cantabria funds, DGESIC (Spain) grant PB97-0220-C02, and the
Spanish Department for Science and Technology grants AYA2000-2111-E and 
AYA2001-1647-C02. 
\end{acknowledgements}

\appendix
\section{approximation to the function $G(q)$}

The function $G(q)$ is defined from Eq. (7) of the main text. For $q >$ 1, we can 
rewrite $G(q)$ as
\begin{equation} 
G(q) = \frac{1}{\sqrt{q}} \int_{-1}^{+1} \frac{dy}{\sqrt{1 - y^2}} 
(1 - y/q)^{-1/2}  ,
\end{equation}
and expand the factor $(1 + x)^{-1/2}$, $x = - y/q$, $-1 < x < 1$. Then we derive
an approximated law
\begin{equation} 
G(q) \simeq \frac{\pi (1155 + 1680 q^2 + 3072 q^4 + 16384 q^6)}{16384 q^{13/2}} ,
\end{equation}
which works very well at $q >$ 2 (relative deviations less than 0.025\%). However, 
the previous expansion does not work so well in the interval 1 $< q \leq$ 2. Taking 
$u = 1/\sqrt{q - y}$ and $dv = dy/\sqrt{1 - y^2}$, $G(q)$ will be equal to
\begin{equation} 
G(q) = \frac{\pi}{2} \left( \frac{1}{\sqrt{q - 1}} + \frac{1}{\sqrt{q + 1}} \right) 
+ K(q)  ,
\end{equation}
with
\begin{equation} 
K(q) = - \frac{1}{2} \int_{-1}^{+1} dy \frac{\arcsin y}{(q - y)^{3/2}} .
\end{equation}
Expanding the function $\arcsin y$, we can approximate $K(q)$ (and thus, $G(q)$) as 
\begin{equation} 
K(q) \simeq \frac{1 + 2 q}{\sqrt{q + 1}} + \frac{1 - 2 q}{\sqrt{q - 1}} + ... .
\end{equation}
At $q \approx 1$ the first approach (based on the expansion of $(1 - y/q)^{-1/2}$)
underestimates the true behaviour, while the new scheme (based on the expansion of
$\arcsin y$) overestimates $G(q)$. Due to this fact, at 1 $< q \leq$ 2, $G(q)$ is
fitted to a law incorporating both trends. The final result is 
\begin{equation} 
G(q) \simeq \frac{\sum_{i=0}^{2} a_iq^i}{q^{5/2}} + 
\frac{\sum_{i=0}^{2} b_iq^i}{\sqrt{q - 1}} + \frac{\sum_{i=0}^{2} 
c_iq^i}{\sqrt{q + 1}}   ,
\end{equation}
where $a_0 = 2.89283135146944$, $a_1 = 0$, $a_2 = 125.8537652305732$, $b_0 = 
-102.1055508185553$, $b_1 = 227.8006617807037$, $b_2 = -125.694998412877$,
$c_0 = -202.3343397590727$, $c_1 = -93.3006092335309$, and $c_2 = 
124.1342775121369$. The approximation (A.6) is very accurate (relative deviations
$\leq$ 0.01\%), even at $q$ values very close to the singularity ($q$ = 1).

For $q <$ 1, it is more difficult to obtain useful approaches to $G(q)$. After some
tests, we considered the combination of a smooth law (i.e., $d_0 + d_1 q + d_2 q^2 + 
...$) and a trend that is singular at $q$ = 1 (i.e., $(e_0 + e_1 q + e_2 q^2 + 
...)/\sqrt{1 - q}$). At $-1 < q < 1$, from the combined law we infer a good fit  
\begin{equation} 
G(q) \simeq \sum_{i=0}^{8} d_iq^i + 
\frac{\sum_{i=0}^{8} e_iq^i}{\sqrt{1 - q}}   ,
\end{equation}
where $d_0 = 353.7491560566996$, $d_1 = -29.22210584958317$, $d_2 = 
-285.8452207779682$, $d_3 = -296.6766633600129$, $d_4 = -77.10649562470328$,
$d_5 = 236.0210288747884$, $d_6 = 329.3059620964406$, $d_7 = -240.8295381320522$,
$d_8 = 18.61151978935351$, $e_0 = -351.1273028283772$, $e_1 = 205.3847185910097$,
$e_2 = 315.1612274462573$, $e_3 = 171.9981477672829$, $e_4 = -95.1705826992738$,
$e_5 = -320.8255310602542$, $e_6 = -243.8876547053204$, $e_7 = 415.2244998079605$,
and $e_8 = -96.7574150920885$. The relative deviations are again less than 0.02\%.
Of course $G(q)$ = 0 at $q \leq -1$.

\end{document}